\documentclass[twocolumn,preprintnumbers,amsmath,amssymb,floatfix,a4paper]{revtex4-1}

\usepackage{graphicx}
\usepackage{dcolumn}
\usepackage{bm}
\graphicspath{{./},{./figs/}}
\usepackage[latin1]{inputenc}

\begin{document}


\title{Conditions for the occurrence of Coulomb blockade in phosphorene quantum dots at room temperature}


\author{H. A. Melo}
\affiliation{Universidade Federal do Cear\'a, Departamento de F\'{i}sica, Caixa Postal 6030, 60440-900 Fortaleza, Cear\'a, Brazil}
\author{M. A. Lino}
\affiliation{Universidade Federal do Piau\'{i}, Departamento de F\'{i}sica, 64049-550 Teresina, Piau\'{i}, Brazil}
\author{D. R. da Costa}
\author{A. Chaves}
\author{J. M. Pereira Jr.}
\author{G. A. Farias}
\author{J. S. de Sousa}
\email{jeanlex@fisica.ufc.br}
\affiliation{Universidade Federal do Cear\'a, Departamento de F\'{i}sica, Caixa Postal 6030, 60440-900 Fortaleza, Cear\'a, Brazil}



\begin{abstract}
We study the addition energy spectra of phosphorene quantum dots focusing on the role of dot size, edges passivation, number of layers and dielectric constant of the substrate where the dots are deposited. We show that for sufficiently low dielectric constants ($\varepsilon_{sub} < 4$), Coulomb blockade can be observed in dot sizes larger than 10 nm, for both passivated and unpassivated edges. For higher dielectric constants (up to $\varepsilon_{sub} = 30$), Coulomb blockade demands smaller dot sizes, but this depends whether the  edges are passivated or not. This dramatic role played by the substrate is expected to impact on the development of application based on phosphorene quantum dots.
\end{abstract}

\maketitle

\pagebreak

\section{Introduction} 

The production of single (few) layers of black phosphorus (BP), also known as phosphorene, attracted much attention from the scientific community because of its physical and chemical properties that are potentially useful for nanoelectronics \cite{castellanos2014,xia2014, ling2015, liu2014}. Phosphorene combines characteristics of traditional direct gap semicondutors and the exciting physics of two-dimensional systems. Different from graphene, few-layer phosphorene has a large band gap, varying between 0.3 eV and 2.0 eV, that can be tuned by the number of stacked layers \cite{tran2014,ref6,ref7,ref8,rudenko2015}. These properties inspired the demonstration of many different applications like field effect transistors \cite{liu2014,ref5}, detectors \cite{cui2015}, modulators \cite{zheng2017} and sensors \cite{kou2014}. The possibility of developing a phosphorene based technology triggered 
a huge number of studies to understand and control its properties. For example, it was recently shown that phosphorene exhibits, depending on the substrate where it is deposited, very large exciton binding energies \cite{chaves2015,zhang2018}. de Sousa \textit{et al.} calculated the exciton fine structure of monolayer phosphorene quantum dots deposited on different substrates. For QDs large enough to reproduce the properties of infinite layer, they demonstrated that difference in the photoluminescence peaks of two independent studies of Zhang (1.67 eV) \cite{zhang2016} and Li (1.73 eV) \cite{li2017} are due to the interaction of carriers in the phosphorene layer with the substrate \cite{desousa2017}. In fact, several fundamental studies have shown that the dielectric surroundings have strong influence in the inter-particle interaction in two-dimensional systems \cite{rodin2014,cudazzo2011,berkelbach2013,latini2015,olsen2016}. This interaction can be tuned to produce Coulomb engineered stacks of two dimensional materials \cite{raja2017}.

A natural direction of the research on phosphorene is the fabrication and investigation of the properties of phosphorene quantum dots (PQDs). It is expected that PQDs exhibit interesting physical and chemical phenomena mixing the characteristics of colloidal quantum dots (e.g. size-dependent quantum confinement and surface functionalization) with the properties of two-dimensional systems. PQDs can be fabricated by wet exfoliation methods which allow reasonable control of the size of QDs. Studies on the fabrication of PQDs reported fairly circular shapes with varying number of layer and sizes ranging between 1 nm and 15 nm, depending on the fabrication method \cite{ref11,ref12,ref13,ref14,vishnoi2018}. In particular, Vishnoi \textit{et al.} \cite{vishnoi2018} reported the production of stable blue-emitting PQDs, and wavelength-dependent photoluminescence (PL). They also reported photoluminescence quenching by electron donors and acceptors. The absence of size dependence in the PL and PL quenching suggests that the QDs surface may have dangling bonds that are partially saturated by these donors/acceptors molecules.

Technological applications of PQDs have also been envisioned and tested. For example, PQDs have been fabricated and employed in ultrafast fiber lasers \cite{du2017}, solar energy conversion \cite{rajbanshi2017} and in non-volatile memories \cite{ref11}. In the latter application, Zhang \textit{ et al.} fabricated low power non-volatile memory and measured the charge storage properties of the device. They reported a high ON/OFF current ratio of the order of $10^4$ at reading voltages of only 0.2 V, which is significantly higher than C$_{60}$ and MoS$_2$ based PVP devices. Although the exact write/erase mechanism of the nonvolatile memories produced by Zhang \textit{et al.} were not described, Lino \textit{et al.} \cite{lino2017} hypotesized that their working principle is very similar to the single (few) electron transistor (SET) device model, and demonstrated that the charging energies of small PQDs are much larger than the thermal energy $k_B T$ and should exhibit Coulomb blockade effects. In particular, Coulomb blockade in two-dimensional materials has only been measured in graphene quantum dots and nanoribbons \cite{sols2007,stampfer2008}, and experimental evidences of Coulomb blockade in phosphorene-based nanostructures are yet to be reported. Constrained by the limitations of the density functional theory (DFT), Lino \textit{et al.} focused on small and isolated PQDs, while in practical applications the dots are deposited in a substrate. However, the substrate-induced dielectric screening between charge carriers in the two-dimensional layer (either QDs or infinite phosphorene sheets) is too strong to be disregarded \cite{desousa2017}. Unfortunately, the calculation of the electronic structure of QDs sitting on dielectric substrates is a challenging task for DFT-based methods.  As the charging energy of a system is determined mainly by the Coulomb interaction among confined particles, new methods to describe for the role of substrate in the addition energy spectra of QDs made of two-dimensional materials are necessary. 

In this work, we calculate the addition energy spectra of PQDs in realistic conditions, focusing on the role of QD size, edges passivation, and substrate, covering a wide range of substrate dielectric constants of materials relevant to nanoelectronics. We demonstrate that these parameters are critical for the observation of Coulomb blockade at room temperature, whereas substrate is more important than dot size, specially in the case where dangling bonds are not passivated. We also map the range of QD sizes and substrate dielectric constants for which Coulomb blockade is expected to occur.

\begin{figure}[ht]
\begin{center}
\includegraphics[width=\columnwidth,clip=true]{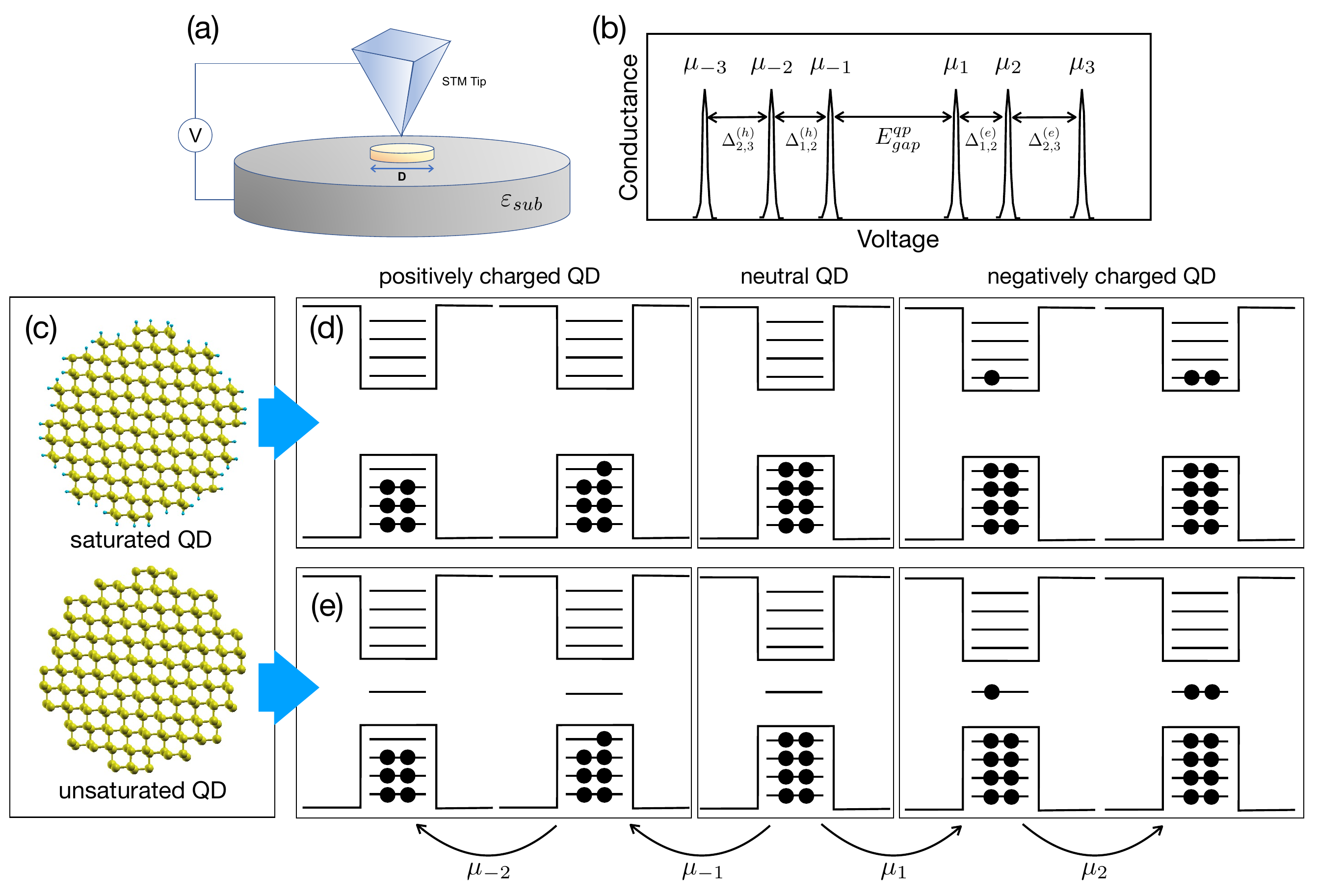}
\caption{\label{fig:qdcharging} Schematics of QD charging spectroscopy. (a) An STM tip is positioned above the QD, and a voltage difference is applied between tip and substrate. (b) Conductance dI/dV as a function of the voltage. The peaks position depends on the charging energies $\mu_n$, and the inter-peaks separation depends on the addition energy spectra $\Delta_{n,n+1}$. (c) QD of 3 nm of diameter with and without saturation of dangling bonds. (d-e) Filling of the single particle states of saturated and unsaturated QDs for several charging states. }
\end{center}
\end{figure}

\section{Methodology}

\subsection{QD charging process and addition energy spectra}

The QD charging process is depicted in Figure \ref{fig:qdcharging}, where it is assumed that after particles are added/removed to/from the QD, the system quickly thermalizes to its ground state, such that the added particles fill unoccupied single particle states according to Hund's rule. Thermal fluctuations are disregarded. The addition/removal of electrons in a neutral QD (also regard as the reference system with $N$ electrons) can be measured by scanning tunneling spectroscopy (STM) \cite{banin1999, cao2014, cheng2017,liang2014}. The conductance $dI/dV$ as a function of the voltage $V$ applied between the STM tip and substrate exhibits peaks whose positions and inter-peaks separations depend on the charging energy and addition energy spectrum of the QD.  As electrons are either added or removed the total energy of the confined system is modified. The charging energy $\mu_n$ is the energy needed to add one electron to a QD already containing $n-1$ electrons. This quantity is calculated as \cite{franceschetti2000a,franceschetti2000b, melnikov2004,oliveira2008,thean2001,he2005}:

\begin{equation}
\label{eq:charging}
\mu_n = E(n)-E(n-1)
\end{equation}

\noindent where $E(n)$ is the total energy of the QD containing $n$ electrons. The addition energy $\Delta_{n,n+1}$ indicates how much more energy is needed to add the $(n+1)_{th}$ electron compared to the energy to add the $n_{th}$ electron. This is given by:

\begin{equation}
\label{eq:addition}
\Delta_{n,n+1}^{(e)} = \mu_{n+1} - \mu_n.
\end{equation}

The above definition can also be used to determine the charging energies for holes. In this sense, $\mu_{-1}$ is the energy to add one hole to the neutral QD. The difference $E_{gap}^{qp} = \mu_1-\mu_{-1}$ is the quasiparticle gap i.e., the energy necessary to remove one electron from the highest occupied orbital $h_1$ and place it in the lowest unoccupied orbital $e_1$ of an identical QD at an infinite distance, such that the electron and hole do not interact \cite{franceschetti2000a,melnikov2004,oliveira2008}. Analogously, the addition energy for holes is defined as $\Delta_{n,n+1}^{(h)} = \mu_{-n} - \mu_{-(n+1)}$.  

The determination of the charging energies of QDs depend on the calculation of the total energy of a system containing $N$ electrons. This can be done by a number of methods like DFT-based \emph{ab initio} methods \cite{desousa2017}, semi-empirical pseudopotential \cite{franceschetti2000a,franceschetti2000b, an2007} and effective mass theory \cite{thean2001}. Here, we first calculated the single particle states of the QDs using a tight-binding (TB) method. These states are then used to construct Slater determinants representing the total wavefunction of the $N$-electrons systems. The total energy of the system is calculated according to Hartree-Fock theory \cite{szabo1996}.  

\subsection{Single-particle electronic structure}

Circular PQDs were formed by generating a large sheet of phosphorene (up to three layers adopting AB stacking) with armchair (zigzag) direction aligned to the $x$ ($y$) axis. The energy spectrum of the PQDs was calculated by solving Schroedinger equation represented in a linear combination of atomic orbital (LCAO) basis, such that the effective Hamiltonian reads $\hat{H}  = \sum_i \epsilon_i |i\rangle\langle i | + \sum_{i,j} t_{i,j} |i\rangle\langle j |$. The generalized index $i = \{\vec{R}_i,\alpha,\nu\}$ represents the orbital $\nu$ of the atomic species $\alpha$ at the atomic site $\vec{R}_i$. $\epsilon_i$ represents the onsite energy of the i-th site, and $t_{i,j}$ represents the hopping parameter between i-th and j-th sites. As for the hopping parameters and lattice constants, we adopted the parameters of Rudenko \textit{et al.} \cite{rudenko2015}, fitted from phosphorene band structure calculations based on state-of-the-art GW method. QDs constructed as described above exhibit interface states due to presence of dangling bonds in the QD borders. Since an experimental method to saturate those dangling bonds may still lack, we will study both unsaturated and saturated QD configurations. In the former case, the extra electrons are added to the mid-gap edge states. In the latter case, the extra electrons are added to the conduction band of the QDs. For this, we ignore the edge states, and assume that the wavefunctions and energy difference between adjacent levels of the lowest few conduction band states of saturated and unsaturated QDs are nearly identical. 

\subsection{Total energy calculation}

For a neutral QD with $N$ electrons, the ground state wavefunctions of a neutral and charged QDs (with either one hole or electron) are given by the following Slater determinants:

\begin{widetext}

\begin{flalign}
\Phi_{N-1}(\vec{r}_1,...,\vec{r}_{N-1}) &= \mathcal{A}[\psi_1(\vec{r}_1),\bar{\psi}_1(\vec{r}_2),...,\psi_{vbm}(\vec{r}_{N-1})] \\
\Phi_{N}(\vec{r}_1,...,\vec{r}_N) &= \mathcal{A}[\psi_1(\vec{r}_1),\bar{\psi}_1(\vec{r}_2),...,\psi_{vbm}(\vec{r}_{N-1}),\bar{\psi}_{vbm}(\vec{r}_N)] \\
\Phi_{N+1}(\vec{r}_1,...,\vec{r}_{N+1}) &= \mathcal{A}[\psi_1(\vec{r}_1),\bar{\psi}_1(\vec{r}_2),...,\psi_{vbm}(\vec{r}_{N-1}),\bar{\psi}_{vbm}(\vec{r}_N), \psi_{cbm}(\vec{r}_{N+1})] \nonumber.
\end{flalign}

\end{widetext}

\noindent where $\psi_i$ ($\bar{\psi}_i$) represents the single particle states with spin up (down) of the QD, $\mathcal{A}$ is the anti-symmetrisation operator, and $vbm$ ($cbm$) stands for valence band maximum (conduction band minimum). In the Hartree-Fock formalism, the total energy of the state  $\Phi_{N}$ is given by \cite{szabo1996}:

\begin{equation}
E_N = \langle\Phi_N | \hat{H}_T(N) | \Phi_N \rangle 
\end{equation}

\noindent where $\langle \Phi_N | \Phi_N \rangle  = 1$, and $\hat{H}_T$ is the Hamiltonian of N interacting electrons:

\begin{equation}
\hat{H}_T(N) = \sum_{i=1}^N \hat{h}_i + \sum_{i=1}^N\sum_{j>i}^N V_{ee}(|\vec{r}_i - \vec{r}_j|), 
\end{equation}

\noindent where $\hat{h}_i  = -(\hbar^2/2m) \nabla_i^2 + V(\vec{r})$ is the single particle Hamiltonian of the i-th electron, and $V_{ee}$ is the electron-electron interaction potential. With the help the Slater-Condon rules to evaluate the expected values of one- and two-body operators action on wavefuctions constructed as Slater determinants \cite{szabo1996}, one can determine close formulae for the addition energies as a function of the number of electrons/holes:

\begin{widetext}

\begin{flalign}
&\mu_{1} = e_1 \\ \nonumber
&\mu_{2} = e_1 + J_{e_1,e_1} \\ \nonumber
&\mu_{3}  = e_2 + 2J_{e_1,e_2} - K_{e_1,e_2} \\ \nonumber
&\mu_{4}  = e_2 + 2J_{e_1,e_2} + J_{e_2,e_2} - K_{e_1,e_2} \\ \nonumber
&\mu_{5}  = e_3 + 2J_{e_1,e_3} + 2J_{e_2,e_3} - K_{e_1,e_3} - K_{e_2,e_3} \\ \nonumber
&\mu_{6}  = e_3 + 2J_{e_1,e_3} + 2J_{e_2,e_3} + J_{e_3,e_3} - K_{e_1,e_3} - K_{e_2,e_3} \\ \nonumber
&\mu_{7}  = e_4 + 2J_{e_1,e_4} + 2J_{e_2,e_4} + 2 J_{e_3,e_4} - K_{e_1,e_4} - K_{e_2,e_4} - K_{e_3,e_4} \\ \nonumber
&\mu_{8} = e_4 + 2J_{e_1,e_4} + 2J_{e_2,e_4} + 2 J_{e_3,e_4} + J_{e_4,e_4} - K_{e_1,e_4} - K_{e_2,e_4} - K_{e_3,e_4} \\ \nonumber
\end{flalign}

\begin{flalign}
&\mu_{-1} = -h_1 \\ \nonumber
&\mu_{-2} = -h_1 - J_{h_1,h_1} \\ \nonumber
&\mu_{-3} = -h_2 - 2J_{h_1,h_2} + K_{h_1,h_2} \\ \nonumber
&\mu_{-4} = -h_2 - 2J_{h_1,h_2} - J_{h_2,h_2} + K_{h_1,h_2} \\ \nonumber
&\mu_{-5}= -h_3 - 2J_{h_1,h_3} - 2J_{h_2,h_3} + K_{h_1,h_2} + K_{h_2,h_3} \\ \nonumber
&\mu_{-6} = -h_3 - 2J_{h_1,h_3} - 2J_{h_2,h_3} - J_{h_3,h_3} + K_{h_1,h_3} + K_{h_2,h_3} \\ \nonumber
&\mu_{-7} = -h_4 - 2J_{h_1,h_4} - 2J_{h_2,h_4} - 2 J_{h_3,h_4} + K_{h_1,h_4} + K_{h_2,h_4} + K_{h_3,h_4} \\ \nonumber
&\mu_{-8} = -h_4 - 2J_{h_1,h_4} - 2J_{h_2,h_4} - 2 J_{h_3,h_4} - J_{h_4,h_4} + K_{h_1,h_4} + K_{h_2,h_4} + K_{h_3,h_4} \\ \nonumber
\end{flalign}

\begin{flalign}
\Delta_{1,2}^{(e)}  =& J_{e_1,e_1} \\  \nonumber
\Delta_{2,3}^{(e)}  =& (e_2 - e_1) + (2 J_{e_1,e_2} - J_{e_1,e_1}) - K_{e_1,e_2} \\ \nonumber
\Delta_{3,4}^{(e)}  =& J_{e_2,e_2} \\ \nonumber
\Delta_{4,5}^{(e)}  =& (e_3 - e_2) + (2 J_{e_1,e_3} + 2 J_{e_2,e_3} - 2 J_{e_1,e_2} - J_{e_2,e_2}) + (K_{e_1,e_2} - K_{e_1,e_3} - K_{e_2,e_3}) \\ \nonumber
\Delta_{5,6}^{(e)}  =& J_{e_3,e_3} \\ \nonumber
\Delta_{6,7}^{(e)}  =& (e_4 - e_3) + (2 J_{e_1,e_4} + 2 J_{e_2,e_4}+ 2 J_{e_3,e_4} - 2 J_{e_1,e_3} - 2 J_{e_2,e_3} - J_{e_3,e_3}) \\ \nonumber & + (K_{e_1,e_3} + K_{e_2,e_3}- K_{e_1,e_4} - K_{e_2,e_4} - K_{e_3,e_4}) \\ \nonumber
\Delta_{7,8}^{(e)}  &= J_{e_4,e_4} \\ \nonumber
\end{flalign}

\begin{flalign}
\Delta_{1,2}^{(h)} =& -J_{h_1,h_1} \\ \nonumber
\Delta_{2,3}^{(h)} =&  - (h_2 - h_1) - (2 J_{h_1,h_2} - J_{h_1,h_1}) + K_{h_1,h_2} \\ \nonumber
\Delta_{3,4}^{(h)} =& -J_{h_2,h_2} \\ \nonumber
\Delta_{4,5}^{(h)} =& -(h_3 - h_2) - (2 J_{h_1,h_3} + 2 J_{h_2,h_3} - 2 J_{h_1,h_2} - J_{h_2,h_2}) - (K_{h_1,h_2} - K_{h_1,h_3} - K_{h_2,h_3}) \\ \nonumber
\Delta_{5,6}^{(h)} =& -J_{h_3,h_3} \\ \nonumber
\Delta_{6,7}^{(h)} =& -(h_4 - h_3) - (2 J_{h_1,h_4} + 2 J_{h_2,h_4}+ 2 J_{h_3,h_4} - 2 J_{h_1,h_3} - 2 J_{h_2,h_3} - J_{h_3,h_3}) \\ \nonumber & - (K_{h_1,h_3} + K_{h_2,h_3}- K_{h_1,h_4} - K_{h_2,h_4} - K_{h_3,h_4}) \\ \nonumber
\Delta_{7,8}^{(h)} =& -J_{e_4,e_4} \\ \nonumber
\end{flalign}

\end{widetext}

The quantities $J_{i,j}$ and $K_{i,j}$ represent the direct Coulomb and exchange energies, respectively:

\begin{eqnarray}
\label{eq:eb} 
J_{i,j}\hspace{-0.05cm}=\hspace{-0.1cm}\int\hspace{-0.25cm}\int\hspace{-0.15cm}\psi_{i}^{*}(\vec{r}_1)\psi_{j}^{*}(\vec{r}_2)V_{ee}(|\vec{r}_1\hspace{-0.1cm}-\hspace{-0.05cm}\vec{r}_2 |) \psi_{i}(\vec{r}_1) \psi_{j}(\vec{r}_2) d\vec{r}_1 d\vec{r}_2, \\
K_{i,j}\hspace{-0.05cm}=\hspace{-0.1cm}\int\hspace{-0.25cm}\int\hspace{-0.15cm}\psi_{i}^{*}(\vec{r}_1)\psi_{j}^{*}(\vec{r}_2)V_{ee}(|\vec{r}_1\hspace{-0.1cm}-\hspace{-0.05cm}\vec{r}_2 |) \psi_{j}(\vec{r}_1) \psi_{i}(\vec{r}_2) d\vec{r}_1 d\vec{r}_2.
\end{eqnarray}

\subsection{Dielectric screening model \label{sec:dielectric}}

The electron-electron interaction potential  $V(|\vec{r}_1-\vec{r}_2 |)$ is given by:
\begin{equation}
\label{eq:dielectric}
V_{ee}(r) = \frac{q^2}{4\pi\varepsilon_0} \frac{\pi}{(1+\varepsilon_{sub})r_0} \left[ H_0\left( \frac{r}{r_0}\right)-Y_0\left( \frac{r}{r_0}\right)\right],
\end{equation}
\noindent where we adopted the model of Rodin \textit{et al.} for the Coulomb interaction between charges confined in a two-dimensional material sandwiched between a substrate with dielectric constant $\varepsilon_{sub}$ and vacuum \cite{rodin2014}. 
$r$ is the distance between particles,  $r_0 = 2\pi \alpha_{2D}/\kappa$, $\kappa = (1+ \varepsilon_{sub})/2$,  $H_0$ and $Y_0$ are the Struve and Neumann functions, and $\alpha_{2D}$ represents the 2D polarizability of the multilayers. This quantity is obtained following the method described by Berkelbach \textit{et al.} \cite{berkelbach2013}, who calculated the real component of static dielectric permittivity $\varepsilon$ as a function of the interlayer distance $d$ of a single phosphorene sheet:

\begin{equation}
\label{eq:eps}
\varepsilon = 1 + \frac{4\pi \alpha_{2D}}{L_z},
\end{equation}

\noindent where $L_z$ is the unit cell size in z direction (perpendicular to the multilayer sheets). $L_z$ is large enough to prevent interaction among BP sheets and their multiple copies imposed by periodic boundary conditions. The dielectric function of multilayer BP sheets was calculated using the Density Functional Theory (DFT) within the Generalized Gradient Approximation (GGA) and norm-conserving Troullier-Martins pseudopotentials, as implemented in SIESTA code \cite{soler96,soler2002}. We used double-zeta basis set (DZP) composed of numerical atomic orbitals of finite range augmented by polarization functions. The fineness of the real-space grid integration was defined by a minimal energy cutoff of 180 Ry. The range of each orbital is determined by an orbital energy confinement of 0.01 Ry. The geometries were considered optimized when the residual force components were less than 0.04 eV/\AA. Due to the well known problem of gap understimation of DFT, we applied the scissors operator such that the single particle gap as function of the number of layers reflected the values obtained by the GW calculations of Rudenko \textit{et al.} \cite{rudenko2015}. The 2D polarizability as function of the number of layers is shown in Table \ref{tab:param}. Our monolayer calculation resulted in $\alpha_{2D} = 4.72$ \AA, which is in good agreement with $\alpha_{2D} = 4.1$ \AA~calculated by Rodin \textit{et al.} \cite{rodin2014}.

\begin{table}[htp]
\caption{\label{tab:param} 2D polarizability of phosphorene sheets with different numbers of layers.}
\begin{center}
\begin{tabular}{cccc}
\hline\hline 
$N$ & $L_z$(\AA) & $Re[\varepsilon(0)]$ & $\alpha_{2D}$(\AA) \\ \hline \hline
1     & 25.26 & 3.35  & 4.72         \\
\multicolumn{3}{c}{} & 4.10\cite{rodin2014} \\ \hline
2     & 25.32 & 6.28  & 10.64       \\
3     & 28.92 & 8.64  & 17.59       \\ \hline\hline 
\end{tabular}
\end{center}
\end{table}%

\subsection{Self-energy correction}

The dielectric discontinuity between the QD and the surrounding materials (vacuum above, and dielectric substrate below) modifies the single particle states $e_i$ and $h_i$ such that they must be corrected to include their polarization self-energy as 

\begin{eqnarray}
e_i \rightarrow e_i + \Sigma_{e_i}^{pol} \\
h_i \rightarrow h_i - \Sigma_{h_i}^{pol}.
\end{eqnarray} 

\noindent The general method to calculate the polarization was described by Fraceschetti et al. \cite{franceschetti2000a,franceschetti2000b} as

\begin{equation}
\Sigma_{\alpha}^{pol} = e \int \psi_\alpha^*(\vec{r}) V_{S}(\vec{r}) \psi_\alpha(\vec{r}) d\vec{r},
\end{equation}

\noindent where $V_S(\vec{r}) = \lim_{\vec{r}' \rightarrow \vec{r}} [G(\vec{r},\vec{r}')-G_{bulk}(\vec{r},\vec{r}')]$, and $G(\vec{r},\vec{r}')$ is the Green's function associated with the system, and $G_{bulk}(\vec{r},\vec{r}')$ is the Green's function of the bulk BP.

\section{Results and discussion}

\subsection{Single-particle states}

\begin{figure}[ht]
\begin{center}
\includegraphics[width=\columnwidth,clip=true]{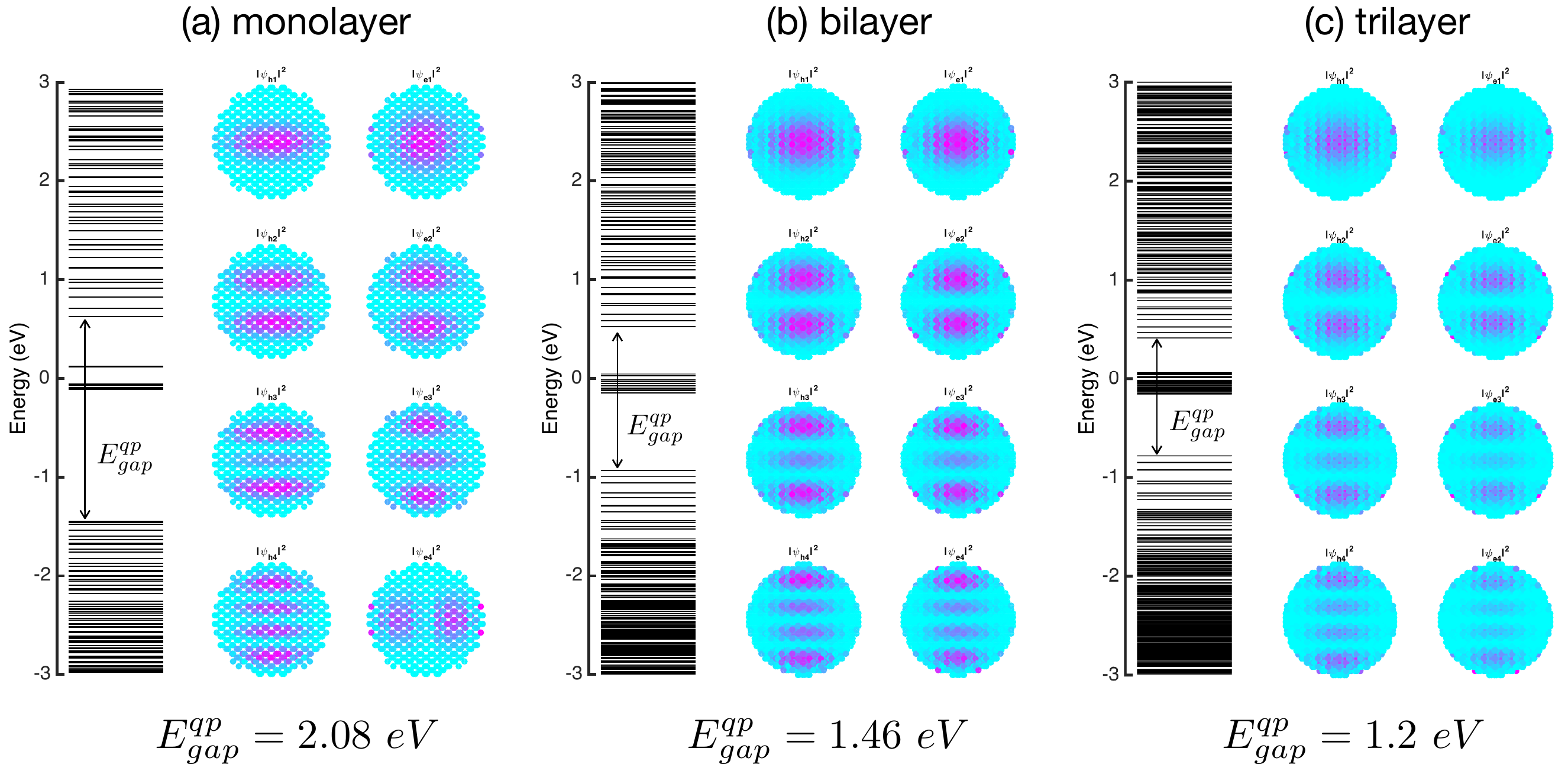}
\caption{\label{fig:spstates} Single particle energy spectra of mono, double and triple layer BPQD with $D$ = 4.5 nm. The squared wavefunctions refer to the four lowest states in conduction and valence bands.}
\end{center}
\end{figure}

The single-particle energy states of mono-, bi- and tri- layer PQDs are shown in Figure \ref{fig:spstates}. The energy spectra changes dramatically with the increase of the number of layers ($N_L$). As in infinite BP layers, the band gap is inversely proportional to $N_L$. Near the conduction band edge, the energy difference between adjacent states also decreases with $N_L$, but the opposite trend is observed near the valence band edge. One interesting feature is the change in the symmetry of the $e_4$ state induced by the stacking of layers. For the monolayer, this state has a node in $x$ direction, while for the bi- and tri-layer this state has three nodes in $y$ direction. This change of symmetry in a consequence of the renormalization of the effective masses caused by the stacking of multiple layers \cite{ameen2016}. Finally, the anisotropy of BP band structure gives rise of anisotropic effective masses. This breaks the radial symmetry of the Hamiltonian, such that the resulting orbitals cannot be labeled according to their angular momentum nomenclature s, p, d, f. 

\begin{figure}[ht]
\begin{center}
\includegraphics[width=\columnwidth,clip=true]{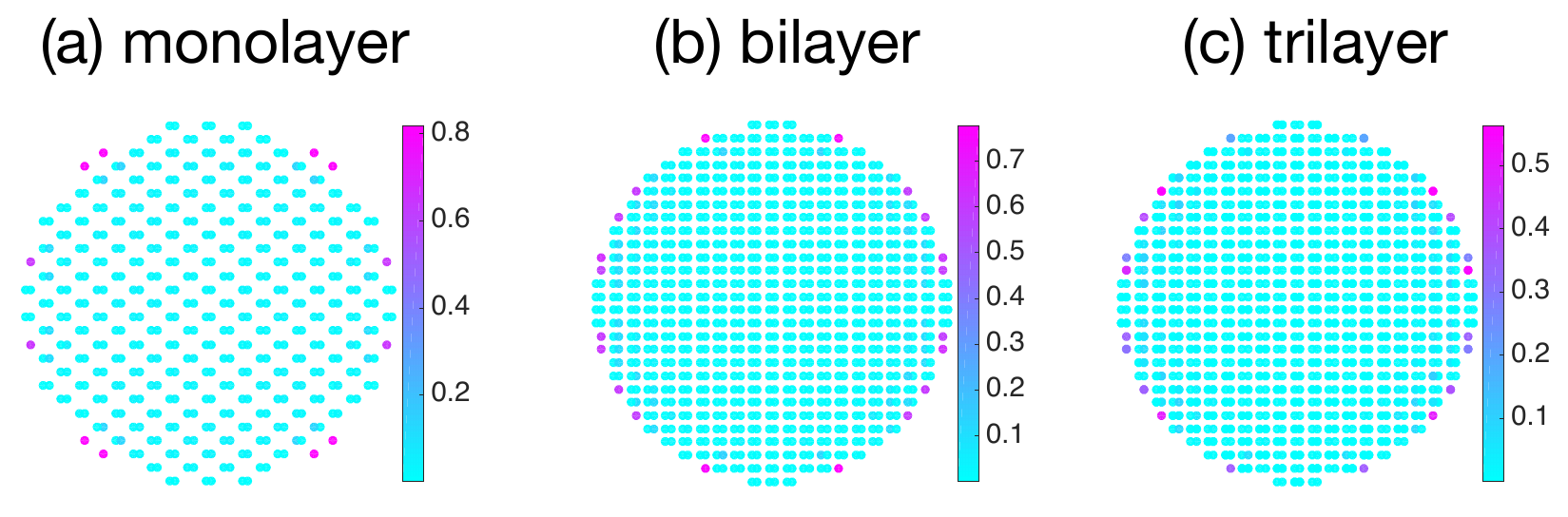}
\caption{\label{fig:defectstates} Linear combination of the squared wavefunctions of the lowest four defect states of mono, double and triple layer BPQD with $D$ = 4.5 nm.}
\end{center}
\end{figure}

The averaged defect wavefunctions of the QDs shown in Figure \ref{fig:spstates} are shown in Figure \ref{fig:defectstates}. This was calculated as the linear combination of the squared wavefunction of the lowest four single particle defect states shown in Figure \ref{fig:spstates}. One can see that electrons confined in defect states are trapped in dangling bonds in the border of the QDs. 

\subsection{Charging energies}

As electrons are added to a QD, they fill unoccupied single particle states. Due to the Coulomb interaction with other confined charges, the total electrostatic energy is also raised. The total energy of charged QDs can be phenomenologically written as 

\begin{equation}
E(n) = \sum n_i (e_i + \Sigma_i^{pol}) + J(n) + K(n), 
\end{equation}

\noindent where $n_i$, $e_i$  and $\Sigma_i^{pol}$ represent the occupancy, energy and polarization self-energy of the $i_{th}$ state, respectively. $J(n)$ and $K(n)$ represent the total electrostatic and exchange energies of the system. The charging energies $\mu_n$, defined as the energy difference between two charging states, still depend on the single particle energy $e_i$ being occupied by the $n_{th}$ electron. The addition energies $\Delta_{n,n+1}^{(e)}$, defined as the difference between two consecutive charging energies, depend only on the energy difference between adjacent single particle states $e_i - e_{i-1}$ ($i$ is the index of single particle states) and on the inter-particle Coulomb $J_{ij}$ and exchange $K_{ij}$ energies between $i_{th}$ and $j_{th}$ quasi-particle states. To understand the charging phenomena, it is instructive to study those individual energetic contributions.

\begin{figure*}[ht]
\begin{center}
\includegraphics[width=\textwidth,clip=true]{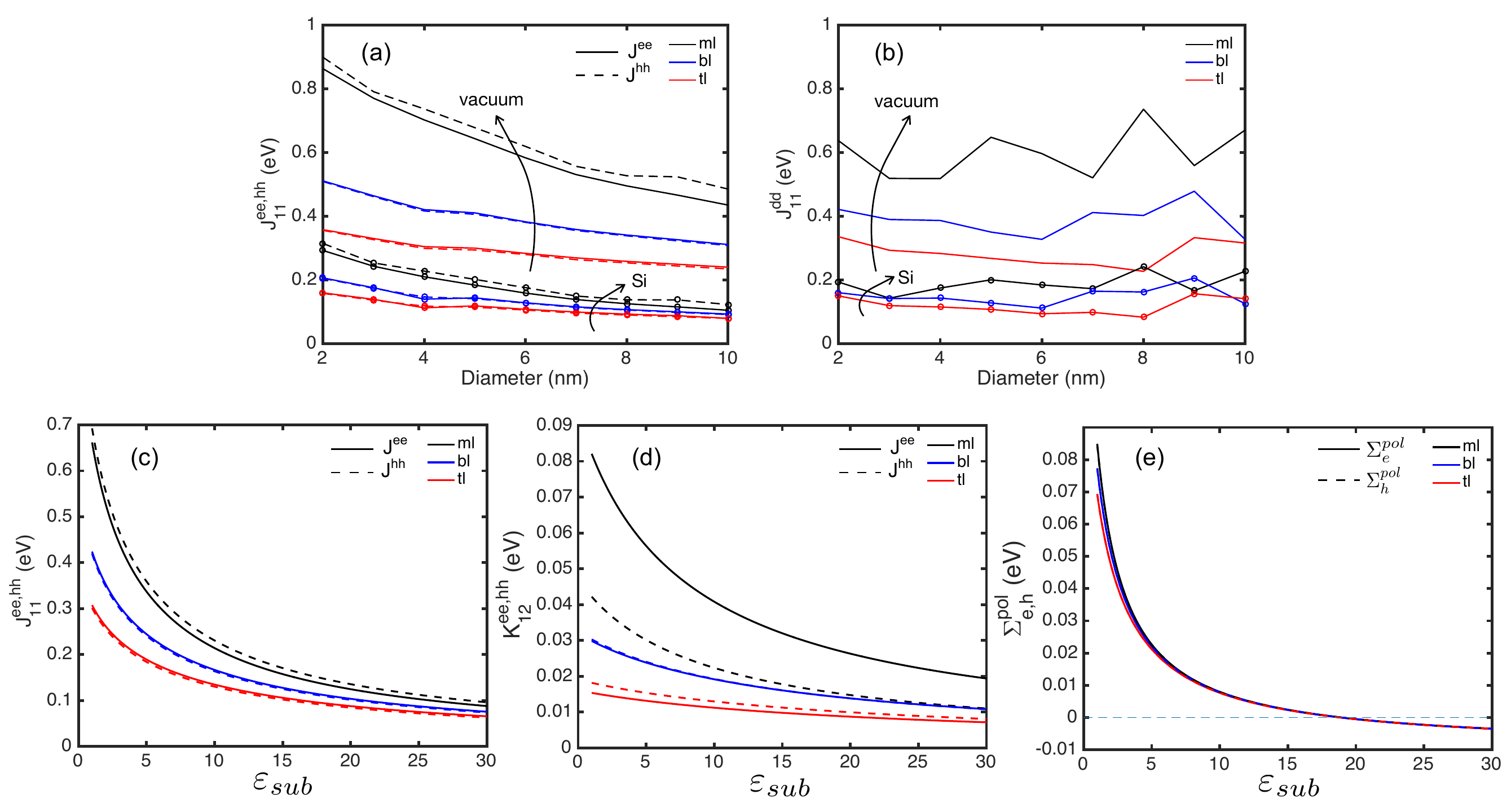}
\caption{\label{fig:coulomb} Size-dependent Coulomb energy between charges (ee, hh, dd) in the ground state of \emph{saturated} (a) and \emph{unsaturated} (b) mono(ml)-, bi(bl)-, and tri(tl)-layer PQDs.  Substrate dependence of the Coulomb (c), exchange (d) and polarization self energies (e) of electrons and holes in a QD with 4.5 nm of diameter.}
\end{center}
\end{figure*}

Figure \ref{fig:coulomb}(a) shows the interaction energy between confined particles (electrons and holes) for saturated multilayer QDs deposited on different substrates. The Coulomb interaction between electrons (holes) $J_{11}^{ee}$ ($J_{11}^{hh}$) is inversely proportional to the QD size and substrate dielectric constant. The repulsion between electrons is weaker than between holes, and this difference between e-e and h-h repulsion decreases with an increase of the number of layers $N_L$. The reason for this can be understood by inspecting the electron and hole ground state wavefunctions (see Figure \ref{fig:spstates}). Due to the interplay between distinct effective masses (and their anisotropy) in each band, the electron wavefunction is spatially distributed in a larger area than the one of holes, making the repulsion between holes stronger than between electrons.  The increase in the number of layers ($N_L$) makes the spatial distribution of the hole wavefunction similar to the electron one. As a consequence, the repulsion between holes becomes nearly identical to the repulsion between electrons in bi- and trilayer QDs. 

Increasing $N_L$ makes more room to accommodate confined charges, resulting in a reduction of the particles repulsion. The ratios between e-e (h-h) repulsion in isolated mono(ml)-, bi(bl)-, and tri(tl)-layer QDs with 5 nm of diameter are $J_{ee}^{bl}/J_{ee}^{ml} = 0.64$ ($J_{hh}^{bl}/J_{hh}^{ml} = 0.60$) and $J_{ee}^{tl}/J_{ee}^{ml} = 0.47$ ($J_{hh}^{tl}/J_{hh}^{ml} = 0.43$). The ratio between repulsion in valence and conduction bands are $J_{hh}^{ml}/J_{ee}^{ml} = 1.05$, $J_{hh}^{bl}/J_{ee}^{bl} = 0.99$ and $J_{hh}^{tl}/J_{ee}^{tl} = 0.98$. In the monolayer, the repulsion between electrons is weaker than the repulsion between holes. As the number of layer increases, the repulsion between electrons become stronger than the repulsion between holes. If those QDs are deposited in a substrate with $\epsilon_{sub} = 11.6$ (Si),  the ratios become $J_{ee}^{bl}/J_{ee}^{ml} = 0.78$ ($J_{hh}^{bl}/J_{hh}^{ml} = 0.70$) and $J_{ee}^{tl}/J_{ee}^{ml} = 0.65$ ($J_{hh}^{tl}/J_{hh}^{ml} = 0.57$), for the bilayer and trilayer, respectively. The ratio between repulsion in valence and conduction bands are $J_{hh}^{ml}/J_{ee}^{ml} = 1.1$, $J_{hh}^{bl}/J_{ee}^{bl} = 0.98$ and $J_{hh}^{tl}/J_{ee}^{tl} = 0.97$. We conclude that the effect of the substrate enhances the inter-particle repulsion in multilayer QDs as compared to the repulsion in the monolayer, but the comparative repulsion of particles in the conduction and valence bands is the same of the isolated QDs. As for the interaction of particles in excited states, we observed the following trend for electrons $J_{11}^{ee}>J_{44}^{ee}>J_{22}^{ee}>J_{33}^{ee}$, and for holes $J_{11}^{hh}>J_{22}^{hh}>J_{33}^{hh}>J_{44}^{hh}$, regardless of QD size and substrate dielectric constant.

It is important to note that the Coulomb interaction in two-dimensional QDs is much higher than in colloidal QDs \cite{franceschetti2000a}. For example, Franceschetti \textit \textit{et al}. studied isolated InAs QDs of different sizes with the semi-empirical pseudopotential method \cite{franceschetti2000b}. They determined that $J_{ee} \approx 0.18$  eV for QDs with diameter of 3 nm. For the same size, we found an interaction four times larger in a monolayer QD, and two times larger for the trilayer QD. 

In unsaturated QDs, additional electrons are trapped in defect states (see Figure \ref{fig:qdcharging}). Thus it is important to study the inter-particle repulsion in those states, which are labeled as $d_n$, where $n$ increases from lower to higher energies. The Coulomb interaction between particles confined in the lowest defect state is shown in Figure \ref{fig:coulomb}(b). As in the case of conduction and valence bands, the repulsion in the defect states reduces as $N_L$ increases. The Coulomb interaction does not exhibit any dependence with QD size, except for some oscillations around a mean value. For isolated QDs, we obtained $\langle J_{11}^{dd} \rangle \approx 0.6$ eV, $\langle J_{11}^{dd} \rangle \approx 0.4$ eV and $\langle J_{11}^{dd} \rangle \approx 0.35$ eV for the mono-, bi- and trilayer QDs, respectively. For QDs deposited on Si, the Coulomb repulsion in mono-, bi- and trilayers becomes nearly identical $\langle J_{11}^{dd} \rangle \leq 0.2$ eV.  The Coulomb interaction between particle confined in upper defect states (not shown here) is numerically comparable and displays similar behavior to $J_{11}^{dd}$. The absence of size dependence in the Coulomb interaction in defects is caused by the fact that the particles are confined in localized states at the edges of the QDs (see Figure \ref{fig:defectstates}), and not on the whole QD area. The oscillations are due to the fact that the distinct spatial arrangement of dangling bonds as the QD size increases. 

The dependence of Coulomb energy on the dielectric constant of the substrate $\varepsilon_{sub}$ of particles occupying the lowest state of the conduction and valence bands of multilayer QDs with 4.5 nm of diameter is shown in Figure \ref{fig:coulomb}(c). The inter-particle repulsion reduces as $\varepsilon_{sub}$ increases. For the monolayer QD, it reduces from 0.7 eV (isolated) to 0.1 when deposited in a substrate with $\varepsilon_{sub} =30$. For trilayer QDs, it reduces from 0.3 eV to 0.06 eV for the same range of dielectric constants. The exchange energy between particles occupying the two lowest states in each band is also shown in Figure \ref{fig:coulomb}(c) but this quantity is approximately 90\% smaller than the Coulomb energy for all cases investigated in this work. Since the main contribution of the addition energies comes from Coulomb energy, Figure \ref{fig:coulomb}(c) shows that the type of substrate is crucial to determine the behavior of $\Delta_{n,n+1}^{(e,h)}$. 


The self energies of $e_1$ and $h_1$ states as function of the dielectric constant of the substrate are shown in Figure \ref{fig:coulomb}(e). $\Sigma_i^{pol}$ is inversely proportional to $\varepsilon_{sub}$, and becomes negative for $\varepsilon_{sub} \ge 18$. $\Sigma_i^{pol}$ is also inversely proportional to $N_L$. But this is a small effect even for substrates with very low dielectric constants. Besides, we obtained that $\Sigma_i^{pol}$ is size- (up to 10 nm of diameter) and state-independent (up to the third excited state in both conduction and valence bands), providing a rigid energy shift for all single particle states. As in the case of InAs QDs, $\Sigma^{pol}$  is nearly identical for electrons and holes \cite{franceschetti2000b}. Interestingly, the polarization self-energies in the InAs QDs are much larger than in PQDs. For isolated QDs  (in vacuum),  $\Sigma^{pol} \approx 0.3$ eV in InAs and $\Sigma^{pol} \approx 0.09$ eV in monolayer PQDs.

The self-energy correction, as described above, results in interesting consequences: it modifies the charging energies, but has negligible effect on the addition energies. It also causes a re-normalization of the quasi-particle gaps of QDs and infinite BP sheets with arbitrary number of layers. In a recent work, de Sousa \textit{et al.} calculated the size-dependent excitonic properties of monolayer PQDs and shown that QD sizes of approximately 10 nm (the largest size considered in that work) display the same properties of infinite monolayer phosphorene sheets \cite{desousa2017}. Due to the fact the $\Sigma^{pol}$ is size-independent, the results shown in Figure \ref{fig:coulomb}(e) suggests that quasi-particle gap of isolated monolayer phosphorene sheets increase by approximately 0.18 eV. The adopted TB parameterization gives a non-polarized quasi-particle gap of $E_{gap}^{qp,0}$ = 1.84 eV for the monolayer sheet \cite{rudenko2015}. The self-energy corrected value becomes $E_{gap}^{qp} = 2.02$ eV. This value is in good agreement with the value of $E_{gap}^{qp}$ = 2.05 eV of monolayer phosphorene measured with STM by Liang \textit{et al.} \cite{liang2014}. This re-normalization depends both on $N_L$ and on the substrate where they are deposited.

\begin{figure}[ht]
\begin{center}
\includegraphics[width=\columnwidth,clip=true]{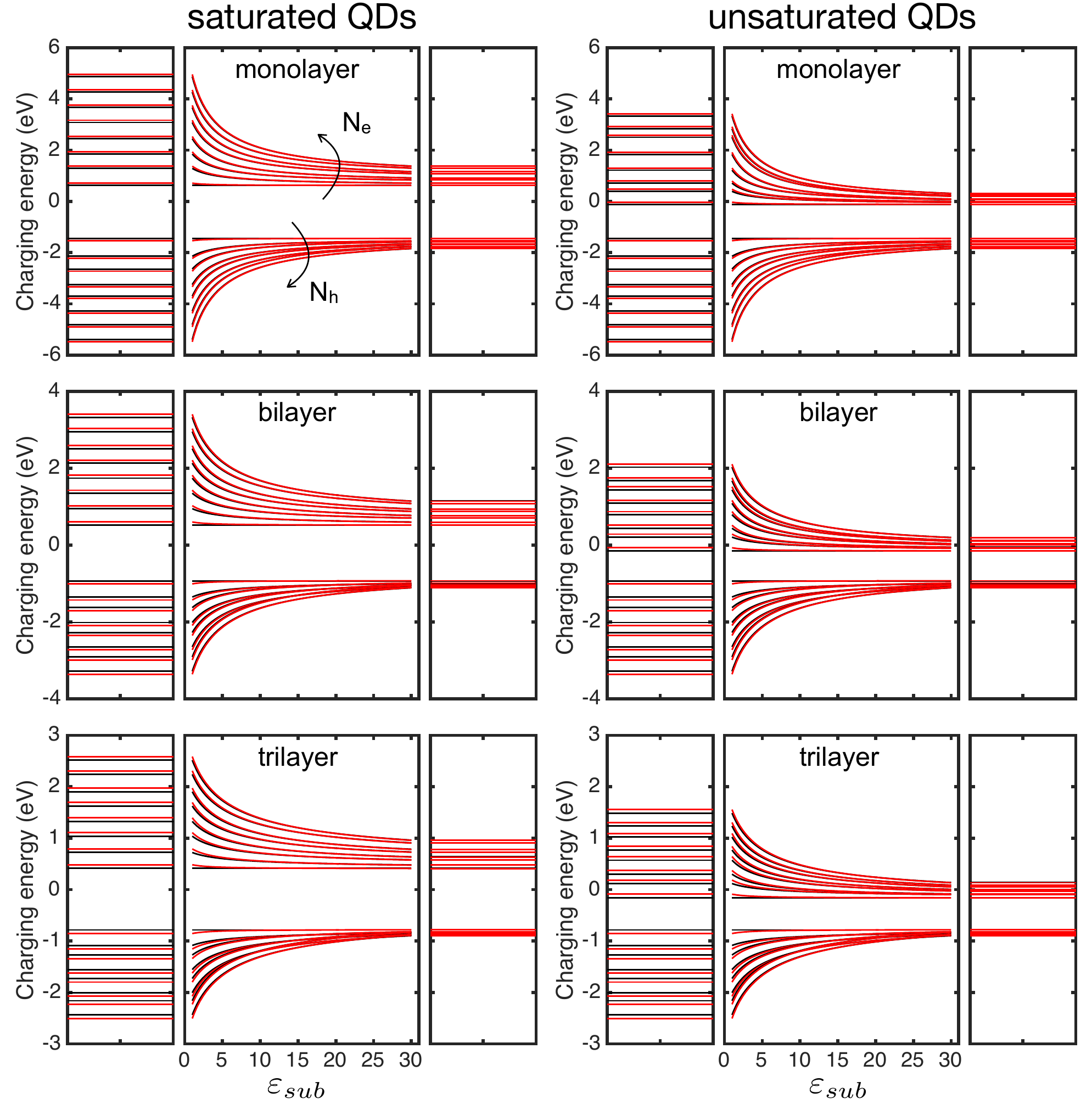}
\caption{\label{fig:charging} Charging energies of a 4.5 nm QD as function of the dielectric constant of the substrate. The arrows show the direction of the increasing number of particles. Black (red) curves represent charging energies without (with) self-energy correction.}
\end{center}
\end{figure}

\begin{figure}[ht]
\begin{center}
\includegraphics[width=\columnwidth,clip=true]{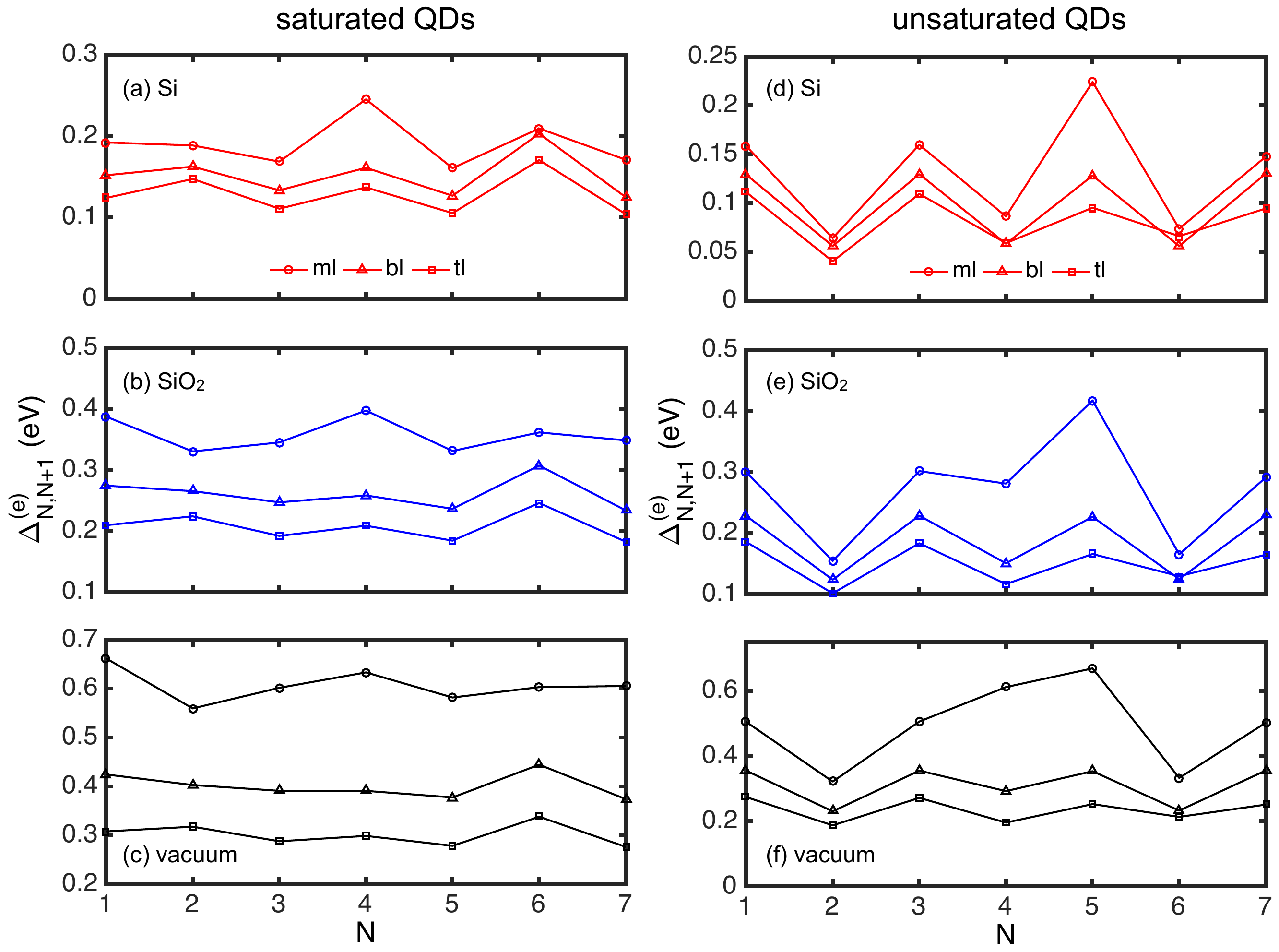}
\caption{\label{fig:addition} Addition energy spectra of saturated and unsaturated QDs of 4.5 nm of diameter deposited on different substrates. }
\end{center}
\end{figure}


Figure \ref{fig:charging} (left panels) shows the charging energy spectra of a saturated QD as a function of the dielectric function of the substrate.  The charging energies strongly depend on a number of factors like QD size, the number of particles confined in the QD, dielectric constant of the substrate, and $N_L$. Self-energy effects are responsible for a small correction of the charging energies. Without this correction, the charging energy of the first electrons would be unaltered, regardless of the dielectric constant of the substrate. Due to existence of mid-gap defects states, the electron charging energies of unsaturated QDs are considerably smaller than for saturated QDs, while the charging energies for holes are equal for both saturated and unsaturated QDs, as shown in the right panels of Figure \ref{fig:charging}. 

The addition energy spectra of QDs with 4.5 nm of diameter in vacuum and deposited in two substrates of technological importance (SiO$_2$ and Si) are shown in Figures \ref{fig:addition}(a)-(c). In vacuum, the average addition energy $\bar{\Delta}= \langle \Delta_{n,n+1} \rangle$ of monolayer QDs is approximately 0.6 eV. When deposited on substrates, $\bar{\Delta}$ reduces to approximately 0.35 eV and 0.2 eV in SiO$_2$ and Si, respectively. The addition energies in multilayer QDs are always smaller than in monolayer ones. The difference between addition energies of bilayer and trilayer QDs is approximately 0.1 eV in vacuum, and it decreases as the dielectric constant of the substate increases. For Si, this difference is less than 0.025eV. This holds for both saturated and unsaturated QDs. For saturated QDs, the addition energies are considerably larger than the thermal energy at room temperature ($k_B T \approx 0.026$ eV), even for substrates with dielectric constants as high as Si, at least for monolayer QDs. For the sake of comparison, we also calculated the charging energies of multilayer QDs as large as 10 nm. The average addition energy $\bar{\Delta}$ of isolated QDs is approximately 0.35 eV (monolayer), 0.26 eV (bilayer) and 0.21 eV (trilayer). For QDs deposited in Si, the average addition energy ranges between  0.07 and 0.08 eV, for all number of layers. 

Figures \ref{fig:addition}(d)-(f) show the addition energy spectra of an unsaturated QD deposited in different substrates. It displays the same general characteristics of the saturated ones. The essential difference is the amplitude of those oscillations. In the saturated QDs, electrons occupy conduction band states that are spatially distributed over the whole QD area, making $\Delta_{n,n+1}$ to depend mainly on the QD size, $N_L$ and type of substrate. In unsaturated QDs, the added electrons are confined in spatially localized defect states. The analytical expressions of the charging energies shows that $\Delta_{m,n}^{(e)} = J_{n/2,n/2}^{dd}$ (for $n = m+1$ and $m$ odd), i.e., when shells are being completely filled. Due to the localized nature of the defects states, the Coulomb repulsion between electrons occupying defect states $J_{mn}^{dd} = J^{dd}$ are nearly identical. The same is true for the exchange energies $K_{mn}^{dd} = K^{dd}$. For unoccupied shells being filled, one has $\Delta_{m,n}^{(e)} \approx (d_n - d_m) + J^{dd} - K^{dd}$ (for $n = m+1$ and $m$ even). If we assume that energy difference between the lowest adjacent states $d_n - d_m = \delta$ is nearly constant, we obtain that the addition energies fluctuate  within the interval $J^{dd}  \leq \Delta_{n,n+1} \leq \delta + J^{dd} - K^{dd}$.  As shown in Figure \ref{fig:coulomb}(b), $J^{dd}$ (and exchange interaction as well) is nearly independent on the QD size. Thus, it is not expected a large difference between addition energies of electrons in unsaturated QDs of different sizes. The same analysis holds for saturated QDs but, oppositely to unsaturated QDs, $\bar{\Delta}$ must exhibit size dependence because all the energetic contributions $\delta$,  $J^{ee}$ and $K^{ee}$ are size-dependent.

\begin{figure}[ht]
\begin{center}
\includegraphics[width=0.9\columnwidth,clip=true]{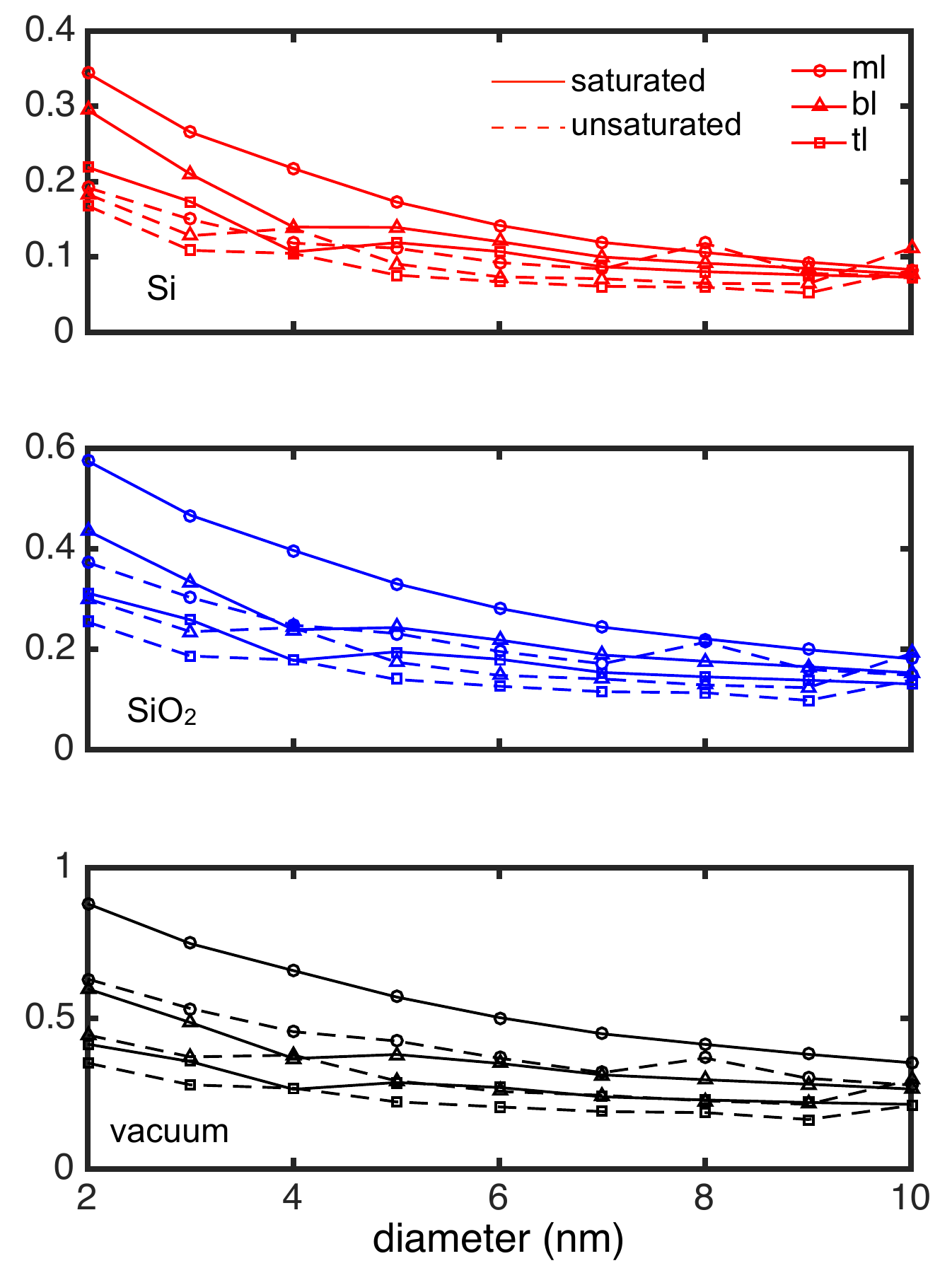}
\caption{\label{fig:sizedep} Size dependence of the average addition energy (up to eight electrons) $\bar{\Delta}$ of multilayer PQDs on differente substrates. }
\end{center}
\end{figure}

The size dependence of $\bar{\Delta}$ for saturated QDs, shown in Figure \ref{fig:sizedep}, follows a scaling law of the type $\bar{\Delta} \propto R^{-\gamma}$ ($\gamma > 0 $), where $R$ is the QD radius. This behavior occurs because all major contributions of $\bar{\Delta}$ are size-dependent. These contributions are (i) the energy difference between adjacent states in the conduction band whose size dependence is $R^{-\alpha}$  ($\alpha \leq 2$, depending on the confinement model), (ii) the Coulomb and (iii) exchange energies whose size dependence is $R^{-\beta}$ ($\beta \leq 1$). $\bar{\Delta}$ is also inversely proportional to the number of layers, and the difference between layers reduces with the QD size and dielectric constant. For an isolated QD with 10 nm of diameter,  $\bar{\Delta}$ is 0.35 eV, 0.26 eV and 0.21 eV for the mono-, bi and trilayer cases, respectively. For the same QD deposited in SiO$_2$, $\bar{\Delta}$ reduces to 0.18 eV, 0.15 eV and 0.13 eV, respectively. For Si as the substrate, $\bar{\Delta}$ becomes as low as 0.08 eV for all cases. 

For unsaturated QDs, $\bar{\Delta}$ is inversely proportional to $N_L$,  and seems to exhibit very weak size dependence for diameters up to 5 nm. For larger sizes, $\bar{\Delta}$ fluctuates around a mean value, without displaying any noticeable size dependence. $\bar{\Delta}$ for unsaturated QDs are smaller than of saturated ones, except for large diameters, where fluctuations may occasionally make it larger than the value of saturated QDs. The absence of size dependence is caused by the localized nature of defect states, for which neither the inter-states energy difference nor the Coulomb and exchange energies between defect states are size-dependent quantities. 

\subsection{Occurrence of Coulomb blockade}

Our calculations show that the addition energies of PQDs is inversely proportional to the number of layers, dielectric constant of the substrate, regardless the saturation state of dandling bonds in the QD borders. Size dependence appears only if dangling bonds are saturated. If the saturation of the dangling bonds is partial, it is expected a mixed behavior between saturated and unsaturated cases. It was also shown that, due to the two-dimensional geometry of the QDs, the Coulomb and exchange interactions are enhanced as compared to the tridimensional QDs, raising the charging energies much above the thermal energy $k_B T$ even at room temperature.

\begin{figure}[ht]
\begin{center}
\includegraphics[width=0.8\columnwidth,clip=true]{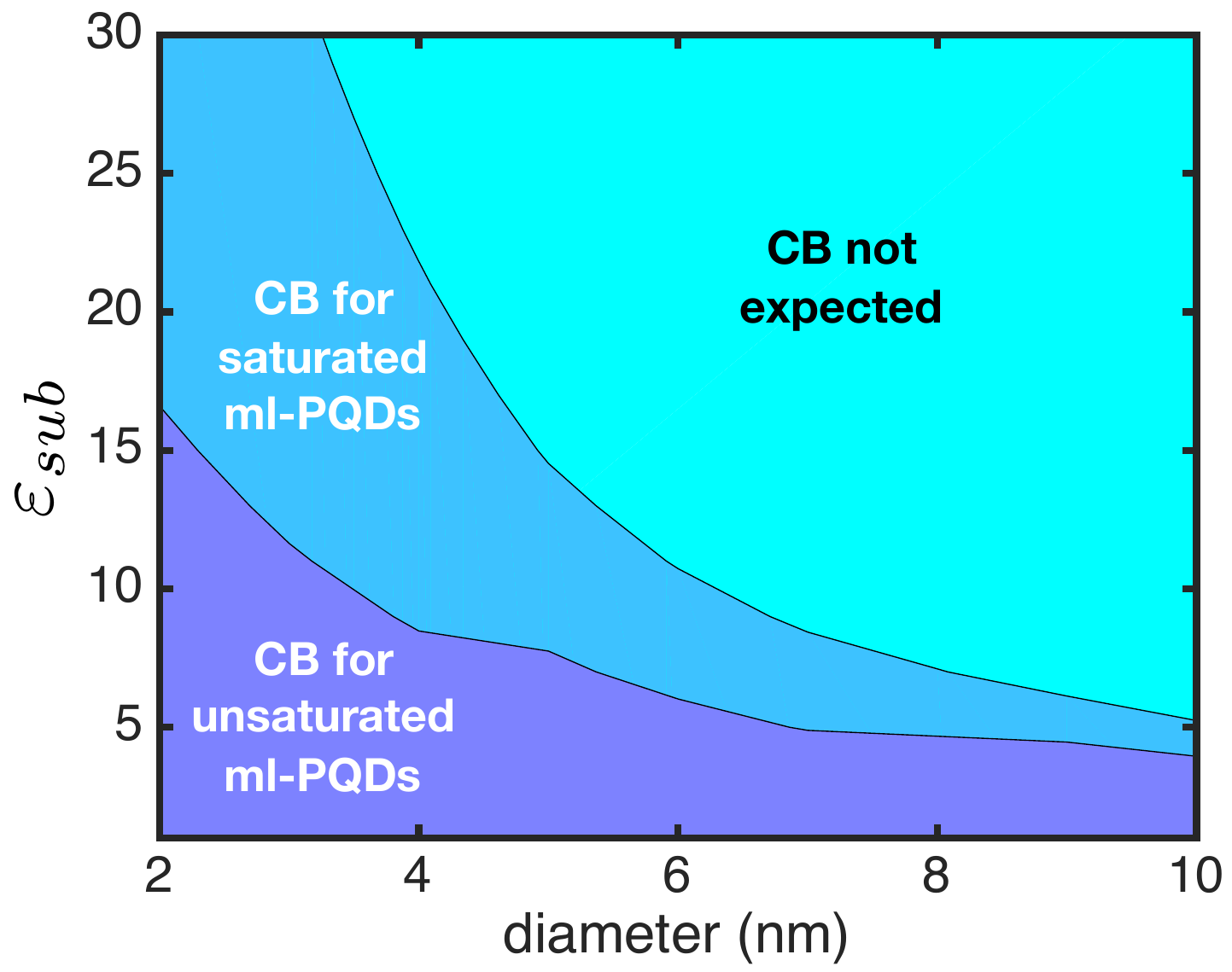}
\caption{\label{fig:diagml} Diagram showing the combination of monolayer PQD sizes and dielectric constants of substrates for which Coulomb blockade at room temperature is expected to occur. The adopted cut-off energy is $\bar{\Delta} = 0.15$ eV.}
\end{center}
\end{figure}

Such high addition energies are evidence that PQDs are good candidates for the development of single electron transistors (SET) whose working principle is governed by the Coulomb blockade (CB) effect. There are two fundamental conditions to observe CB in the electrical transport through small quantum islands: (i) the single electron addition energy must be much larger than the thermal energy, (ii) the tunnel coupling between the quantum islands and the leads has to be small to ensure a long lifetime $\Delta t$ of the electrons in the quantum island such that the uncertainty in energy $\Delta E \approx \hbar/ \Delta t$ must not exceed the addition energy. The former depends only on the electronic properties of the quantum islands, while the later depends on the design of the device. If we consider that $6 k_B T \approx 0.15$ eV is a safe cut-off energy above which CB may occur at room temperature, our results show that obtaining $\bar{\Delta} \approx 0.15$ eV largely depends on several factors. The substrate seems to be more relevant than QD size, depending on the degree of passivation of the dangling bonds. As shown in Figure \ref{fig:sizedep}, if the substrate has low enough dielectric constant ($\varepsilon_{sub} \leq \varepsilon_{SiO_2}$), one can observe Coulomb blockade in QDs as large 10 nm with up to three layers. We summarize the combination of monolayer QD sizes and substrate dielectric constants for which CB is expected to occur with the diagram shown in Figure \ref{fig:diagml}. For tiny saturated PQDs up to 3 nm, CB is expected for substrates with dielectric constant up to $\varepsilon_{sub} = 30$. For unsaturated PQDs, this range of $\varepsilon_{sub}$ is much narrower. If $\varepsilon_{sub}$ is low enough, CB can be observed in QDs with sizes larger than 10 nm for both saturated and unsaturated cases. 

\subsection{Comparison with DFT}

The calculation of the addition energy spectra of QDs deposited on substrates is a challenging task for DFT based methods, and approximated methods like the one presented in this manuscript is a promising alternative to calculated the effect of substrates.  Anyhow, it is expected that the two approaches agree for the case where QDs are in vacuum. In order to perform this comparison, it is important to understand and minimize the differences between the present method and what can be done within DFT framework. 

In a previous work, Lino \textit{et al.} calculated the addition energy spectra of small PQDs with DFT \cite{lino2017}. They saturated the dangling bonds in the QD edges with hydrogen atoms. They also performed a geometry optimization step after the addition of each electron. The saturation of dangling bonds enlarges a little bit the QD sizes as compared to the unsaturated case. Complicated methods to eliminate defects states within TB formalism have been proposed, but they are either complicated or not reliable. Geometry optimization changes the distance between atoms, compared to the initial atomic positions. However, the adopted TB scheme were parameterized for fixed interatomic distances \cite{rudenko2015}. Thus, in order to compare the results obtained with the present HF based method with the ones obtained with DFT, we eliminate from DFT calculations all features that could not be reproduced by our method: saturation of dangling bonds and geometry optimization. 

\begin{figure}[ht]
\begin{center}
\includegraphics[width=0.9\columnwidth,clip=true]{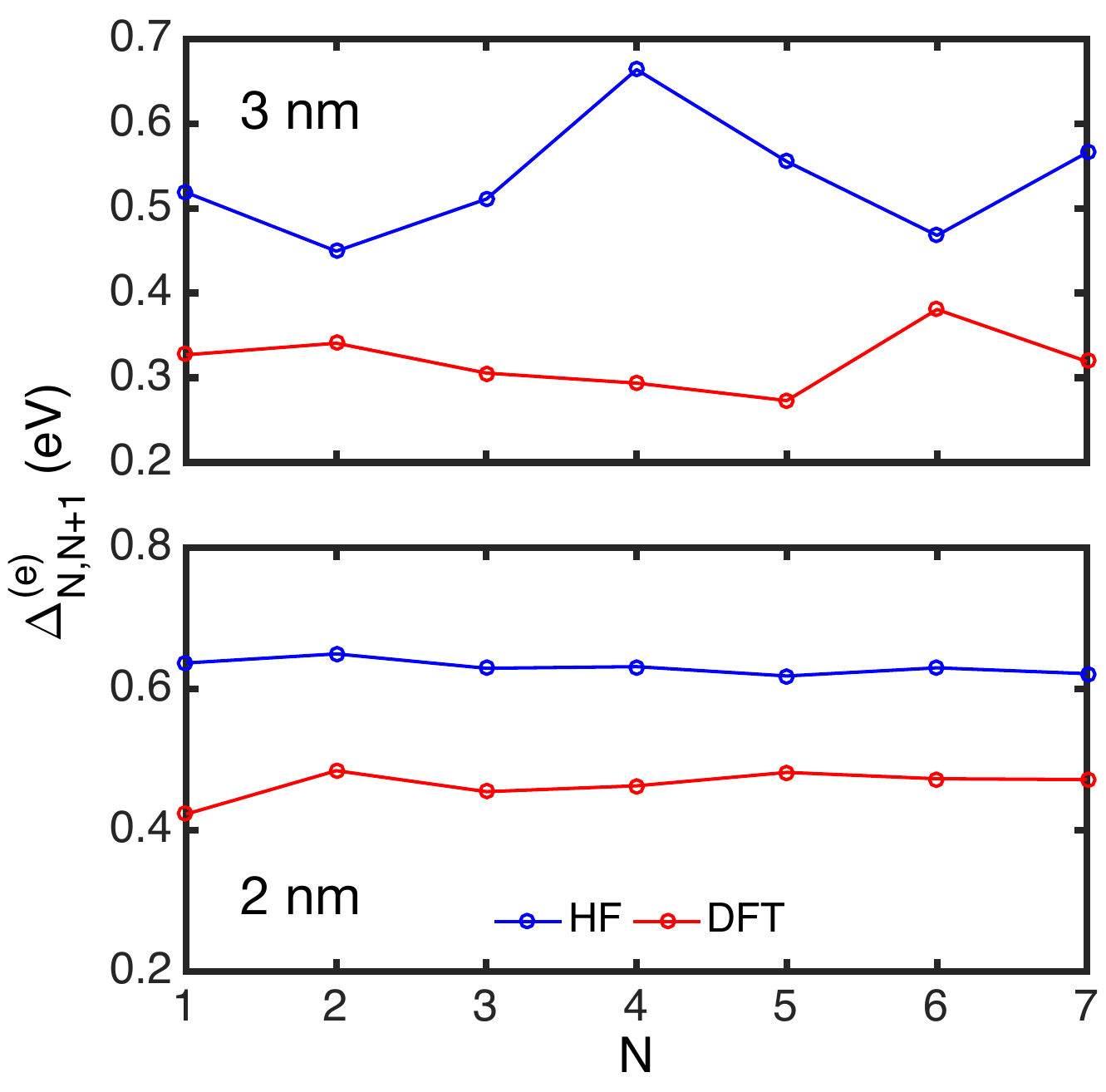}
\caption{\label{fig:comparison} Comparison of the addition energy spectra of unsaturated monolayer QDs with 2 m and 3 nm of diameter. }
\end{center}
\end{figure}

Figure \ref{fig:comparison} compares the addition energy spectra, calculated with DFT and HF methods, of unsaturated monolayer QDs. The DFT calculations were performed as described in Section \ref{sec:dielectric}. The results of both methods exhibit a general resemblance, with the HF-based method providing addition energies that are, in average, 0.2 eV larger than the DFT one. As a general behavior, the addition energies remains fluctuating around a mean value, as the number of confined electrons increases. The HF-computed average addition energies $\langle \Delta_{N,N+1}^{(e)} \rangle$ is 0.63 eV and 0.53 eV for QDs with 2 nm and 3 nm, respectively.  The DFT-computed mean addition energies $\langle \Delta_{N,N+1}^{(e)} \rangle$ is 0.46 eV and 0.32 eV for QDs with 2 nm and 3 nm, respectively. 

The difference between calculations are explained as follows. In one hand, DFT includes correlation effects in the self-consistent solution of Kohn-Sham equation, while our non self-consistent HF method does include correlation effects at all. Correlation adjusts the total electron density to accommodate inter-particle repulsion, reducing the Hartree energy component of the total energy. On the other hand, DFT is known to severely underestimate quasi-particle energies. For example, it is known that DFT-GGA-PBE underestimates the band gap of single layer phosphorene by more than 1 eV as compared to $G_0W_0$ approximation \cite{tran2014}.

In both methods, the charging and addition energies were calculated using Equations \ref{eq:charging} and \ref{eq:addition}, respectively. Thus, any difference between methods completely depends on how the total energies are calculated. We remind that the quasi-particle gap $E_{gap}^{qp}$ of $N$ electrons systems is defined as the difference between the ionization energy $I_N = E(N-1) - E(N)$ and the electron affinity $A_N = E(N) - E(N+1)$ \cite{melnikov2004}, such that $E_{gap}^{qp} = I_N - A_N = E(N+1) - 2E(N) + E(N-1)$. This expression has the same structure of the addition energies $\Delta_{N,N+1}^{(e)}$ in Equation \ref{eq:addition}, where the reference number of electrons $N$ increase one by one to mimic the charging process. Thus, $\Delta_{N,N+1}^{(e)}$ essentially calculates quasi-particle gaps of negatively charged systems with a increasing number of electrons. 

It is well known that even when DFT-GGA act as a good approximation of ground state properties, it underestimates $I_N$ and overestimates $A_N$ by approximately $\Sigma_N/2$ \cite{kronik2012}, such that we have

\begin{equation}
\label{eq:quasiparticle}
E_{gap,N}^{qp}  = E_{gap,N}^{KS} + \Sigma_N,
\end{equation}

\noindent where $E_{gap,N}^{KS} = e_{L}^{KS}(N)  - e_{H}^{KS}(N)$ is the Kohn-Sham band gap of the reference system with $N$ electrons, $e_{L}^{KS}(N)$ and $e_{H}^{KS}(N)$ represent the energies of the lowest unoccupied and highest occupied molecular orbitals as calculated by Kohn-Sham (KS) equation, and $\Sigma_N$ is the self-energy correction. $\Sigma_N$ can also be regarded as a measure the finite variation of the exchange-correlation potential $v_{xc}(\vec{r})$ extended everywhere in the solid due to an infinitesimal variation of the density $n(\vec{r})$:

\begin{equation}
\Sigma_N = \left( \frac{\delta E_{xc}[n]}{\delta n(\vec{r})}\|_{N+1} - \frac{\delta E_{xc}[n]}{\delta n(\vec{r})}\|_{N} \right) + O\left( \frac{1}{N} \right).
\end{equation}

It has been shown that $\Sigma_N$ can be of the order of 1.0 eV for small Si$_{1-x}$Ge$_x$ nanocrystals \cite{oliveira2008}. For mono-, bi- and trilayer phosphorene, $\Sigma$ is of the order of 1.1 eV, 0.8 eV and 0.8 eV, respectively \cite{tran2014}. More sophisticated approaches to deal with exchange and correlation, eg. solving Dyson equation within GW approximation, allows us to calculate the actual quasiparticle gaps of systems directly from the KS gap. This is where the advantage of our method stands out. Our adopted TB scheme was parameterized from a state-of-the-art GW corrected band structure \cite{rudenko2015}. Thus, our HF-based quasiparticle gap of the reference system with $N$ electrons (neutral system) embed the $\Sigma_N$ correction that is underestimated by our DFT-GGA calculation. This also holds for charged systems. The above considerations explain why our HF-based calculation provides larger of the addition energies as compared to the DFT-based calculation, as shown Figure \ref{fig:comparison}.

\section{Conclusions}

In conclusion, we calculated the addition energy spectra of multilayer PQDs for wide range of dot sizes and dielectric constants of the substrates where they are deposited. We also investigated the role of edges passivation on the addition energy spectra. We consistently obtained addition energies higher than the thermal energy $k_BT$. This suggests that Coulomb blockade at room temperature can be observed in PQDs, depending on trade-off between dot size, dielectric constante of the substrate and passivation state of the QD edges: the larger the dot size, the smaller is the dielectric constant of the substrate that allows for CB at room temperature. On the other hand, observing CB in smaller dots depends on the passivation state of the edges. If the edges are fully passivated, CB is observed for any substrate with $\varepsilon_{sub}$ up to 30. If edges are unpassivated, CB can only be observed for $\varepsilon_{sub}$ up to 15. This dramatic role of the substrate is expected to impact not only in the development of charge storage applications of PQDs, but also in optical applications, where dielectric screening effects plays a major role. Finally, we emphasize that the advantage of our methodology goes beyond simplicity of implementation. It allows to predict accurate values of the charging energies of PQDs. Our predictions can be tested experimentally with well established methods like scanning tunneling spectroscopy \cite{banin1999, cao2014, cheng2017}.

\noindent\textbf{Acknowledgements} The authors acknowledge the financial support from the brazilian agencies CNPq and CAPES.



\begin{thebibliography}{0}

\bibitem{castellanos2014} A. Castellanos-Gomez, L. Vicarelli, E. Prada, J. O. Island, K. L. Narasimha-Acharya, S. I. Blanter, D. J. Groenendijk, M. Buscema, G. A. Steele, J. V. Alvarez, H. W. Zandbergen, J. J. Palacios, H. S. J. vander Zant, Isolation and characterization of few-layer black phosphorus, 2D Materials \textbf{1}, 025001 (2014).

\bibitem{xia2014} F. Xia, H. Wang H, Y. Jia, Rediscovering black phosphorus as an anisotropic layered material for optoelectronics and electronics, Nat Commun. \textbf{5}, 4458 Y (2014).

\bibitem{ling2015} X. Ling, H. Wang, S. Huang, F. Xia, M. S. Dresselhaus, The renaissance of black phosphorus, Proc. Nat. Acad. Sci. \textbf{112}, 4523 (2015).

\bibitem{liu2014} H. Liu, A. T. Neal, Z. Zhu, Z. Luo, X. Xu, D. Tom\'anek, P. D. Ye, Phosphorene: An unexplored 2D semiconductor with a high hole mobility. ACS Nano \textbf{8}, 4033 (2014).

\bibitem{tran2014} V. Tran, R. Soklaski, Y. Liang, L. Yang, Layer-controlled band gap and anisotropic excitons in few-layer black phosphorus, Phys. Rev. B \textbf{89}, 235319 (2014).

\bibitem{ref6} J. Sun, H.-W. Lee, M. Pasta, H. Yuan, G. Zheng, Y. Sun, Y. Li, and Y. Cui, A phosphorene-graphene hybrid material as a high-capacity anode for sodium-ion batteries, Nat. Nanotechnol. \textbf{10}, 980 (2015).
\bibitem{ref7} L. Kou, C. Chen, and S. C. Smith, Phosphorene: Fabrication, Properties, and Applications, J. Phys. Chem. Lett. \textbf{6}, 2794 (2015).
\bibitem{ref8} J. Dai and X. C. Zeng, Bilayer Phosphorene: Effect of Stacking Order on Bandgap and Its Potential Applications in Thin-Film Solar Cells, J. Phys. Chem. Lett. \textbf{5}, 1289 (2014).

\bibitem{rudenko2015} A. N. Rudenko, S. Yuan, and M. I. Katsnelson, Toward a realistic description of multilayer black phosphorus: From GW approximation to large-scale tight-binding simulations, Phys. Rev. B \textbf{92}, 085419 (2015).

\bibitem{ref5} L. Li, Y. Yu, G. J. Ye, Q. Ge, X. Ou, H. Wu, D. Feng, X. H. Chen, and Y. Zhang, Black phosphorus field-effect transistors, Nat. Nanotechnol. \textit{9}, 372, (2014).

\bibitem{cui2015} S. Cui, H. Pu, S. A. Wells, Z. Wen, S. Mao, J. Chang, M. C. Hersam, J. Chen, Ultrahigh sensitivity and layer-dependent sensing performance of phosphorene-based gas sensors, Nat. Commun. \textbf{6}, 8632 (2015).

\bibitem{zheng2017} J. Zheng,  X. Tang,  Z. Yang,  Z. Liang,  Y. Chen,  K. Wang,  Y. Song, Y. Zhang, J. Ji,  Y. Liu,  D. Fan,  H. Zhang, Optical Modulation: Few Layer Phosphorene Decorated Microfiber for All Optical Thresholding and Optical Modulation, Adv. Opt. Mat \textbf{5}, 1700026 (2017).

\bibitem{kou2014} L. Kou, T. Frauenheim, C. Chen, Phosphorene as a Superior Gas Sensor: Selective Adsorption and Distinct I-V Response., J. Phys Chem Lett. \textbf{5}, 2675 (2014).

\bibitem{chaves2015} A. Chaves, T. Low, P. Avouris, D. \c{C}akir, and F. M. Peeters, Anisotropic exciton Stark shift in black phosphorus, Phys. Rev. B \textbf{91}, 155311 (2015).

\bibitem{zhang2018} G. Zhang, A. Chaves, S. Huang, F. Wang, Q. Xing, T. Low, H. Yan, Determination of layer-dependent exciton binding energies in few-layer black phosphorus, Science Advances \textbf{4}, eaap9977 (2018).

\bibitem{zhang2016} G. Zhang, A. Chaves, S. Huang, C. Song, T. Low, and H. Yan, Infrared fingerprints of few-layer black phosphorus, Nat. Comm. \textbf{8}, 14071 (2017).

\bibitem{li2017} L. Li,	 J. Kim, C. Jin, G. J. Ye, D. Y. Qiu, F. H. da Jornada, Z. Shi, L. Chen, Z. Zhang, F. Yang, K. Watanabe, T. Taniguchi, W. Ren, S. G. Louie, X. H. Chen, Y. Zhang, and F. Wang, Direct observation of the layer-dependent electronic structure in phosphorene, Nat. Nanotech. \textbf{12}, 21 (2017).


\bibitem{desousa2017} J. S. de Sousa, M. A. Lino, D. R. da Costa, A. Chaves, J. M. Pereira, Jr., G. A. Farias, Substrate effects on the exciton fine structure of black phosphorus quantum dots, Phys. Rev. B \textbf{96}, 035122 (2017).

\bibitem{rodin2014} A. S. Rodin, A. Carvalho, and A. H. Castro Neto, Excitons in anisotropic two-dimensional semiconducting crystals, Phys. Rev. B \textbf{90}, 075429 (2014).

\bibitem{cudazzo2011} P. Cudazzo, I. V. Tokatly, and A. Rubio, Dielectric screening in two-dimensional insulators: Implications for excitonic and impurity states in graphane, Phys. Rev. B \textbf{84}, 085406 (2011).

\bibitem{berkelbach2013} T. C. Berkelbach, M. S. Hybertsen, and D. R. Reichman, Theory of neutral and charged excitons in monolayer transition metal dichalcogenides, Phys. Rev. B \textbf{88}, 045318 (2013).

\bibitem{latini2015} S. Latini, T. Olsen, and K. S. Thygesen, Excitons in van der Waals heterostructures: The important role of dielectric screening, Phys. Rev. B \textbf{92}, 245123 (2015).

\bibitem{olsen2016} T. Olsen, S. Latini, F. Rasmussen, K. S. Thygesen, Simple Screened Hydrogen Model of Excitons in Two-Dimensional Materials, Phys. Rev. Lett. \textbf{116}, 056401 (2016).

\bibitem{raja2017} A. Raja, A. Chaves, J. Yu, G. Arefe, H. M. Hill, A. F. Rigosi, T. C. Berkelbach, P. Nagler, C. Sch\"uller, T. Korn, C. Nuckolls, J. Hone, L. E. Brus, T. F. Heinz, D. R. Reichman, A. Chernikov, Coulomb engineering of the bandgap and excitons in two-dimensional materials,
Nat. Commun. \textbf{8}, 15251 (2017).

\bibitem{ref11} X. Zhang, H. Xie, Z. Liu, C. Tan, Z. Luo, H. Li, J. Lin, L. Sun, W. Chen, Z. Xu, L. Xie, W. Huang, and H. Zhang, Black Phosphorus Quantum Dots, Angew. Chem. Int. Ed. \textbf{54}, 3653 (2015).

\bibitem{ref12} Z. Sofer, D. Bousa, J. Luxa, V. Mazanek, and M. Pumera, Few-layer black phosphorus nanoparticles, Chem. Comm. \textbf{52}, 1563 (2016).

\bibitem{ref13} Z. Sun, H. Xie, S. Tang, X.-F. Yu, Z. Guo, J. Shao, H. Zhang, H. Huang, H. Wang, and P. K. Chu, Ultrasmall Black Phosphorus Quantum Dots: Synthesis and Use as Photothermal Agents, Angew. Chem. Int. Ed. \textbf{54}, 11526 (2015).

\bibitem{ref14} Y. Xu, Z. Wang, Z. Guo, H. Huang, Q. Xiao, H. Zhang, and X.-F. Yu, Quantum Dots: Solvothermal Synthesis and Ultrafast Photonics of Black Phosphorus Quantum Dots, Adv. Optical Mater. \textbf{4}, 1223, (2016).

\bibitem{vishnoi2018} P. Vishnoi, M. Mazumder, M. Barua, S. K.Pati, C. N. R. Rao, Phosphorene quantum dots, Chem. Phys. Lett. \textbf{699}, 223 (2018).

\bibitem{du2017} J. Du, M. Zhang, Z. Guo, J. Chen, X. Zhu, G. Hu, P. Peng, Z. Zheng, H. Zhang, Phosphorene quantum dot saturable absorbers for ultrafast fiber lasers, Sci. Rep. \textbf{7}, 42357 (2017).

\bibitem{rajbanshi2017} B. Rajbanshi, M. Kar, P. Sarkar, P. Sarkar, Phosphorene quantum dot-fullerene nanocomposites for solar energy conversion: An unexplored inorganic-organic nanohybrid with novel photovoltaic properties, Chem. Phys. Lett. \textbf{685}, 16 (2017).


\bibitem{lino2017} M. A. Lino, J. S. de Sousa, D. R. da Costa, A. Chaves, J. M. Pereira, G. A. Farias, Charging energy spectrum of black phosphorus quantum dots, J. Phys. D: Appl. Phys. \textbf{50}, 305103 (2017).

\bibitem{sols2007} F. Sols, F. Guinea, A. H. Castro Neto, Coulomb Blockade in Graphene Nanoribbons, Phys. Rev. Lett. \textbf{99}, 166803 (2007).

\bibitem{stampfer2008} C. Stampfer, J. G\"uttinger, F. Molitor, D. Graf, T. Ihn, K. Ensslin, Tunable Coulomb blockade in nanostructured graphene, Appl. Phys. Lett. \textbf{92}, 012102 (2008).

\bibitem{banin1999} U. Banin, Y. Cao, D. Katz, O. Millo, dentification of atomic-like electronic states in indium arsenide nanocrystal quantum dots, Nature \textbf{400}, 542 (1999).

\bibitem{cao2014} S. Cao, J. Tang, Y. Gao, Y. Sun, K. Qiu, Y. Zhao, M. He, J.-A. Shi, L. Gu, D. A. Williams, W. Sheng, K. Jin,  X. Xu, Longitudinal wave function control in single quantum dots with an applied magnetic field, Sci. Rep. \textbf{5}, 8041 (2014).

\bibitem{cheng2017} J.-Y. Cheng, B. L. Fisher, N. P. Guisinger, C. M. Lilley, Atomically manufactured nickel-silicon quantum dots displaying robust resonant tunneling and negative differential resistance, npj Quantum Materials \textbf{2}, 25 (2017).

\bibitem{liang2014} L. Liang, J. Wang, W. Lin, B. G. Sumpter, V. Meunier, M. Pan, Electronic Bandgap and Edge Reconstruction in Phosphorene Materials, Nano Lett. \textbf{14}, 6400 (2014).

\bibitem{franceschetti2000a} A. Franceschetti, A. Zunger, Addition energies and quasiparticle gap of CdSe nanocrystals, Appl. Phys. Lett. \textbf{76}, 1731 (2000).

\bibitem{franceschetti2000b} A. Franceschetti, A. Williamson, A. Zunger, Addition Spectra of Quantum Dots: the Role of Dielectric Mismatch, J. Phys. Chem. B \textbf{104}, 3398 (2000).

\bibitem{melnikov2004} D. V. Melnikov, J. R. Chelikowsky, Electron affinities and ionization energies in Si and Ge nanocrystals, Phys. Rev. B \textbf{69}, 113305 (2004).

\bibitem{oliveira2008} E. L. de Oliveira, E. L. Albuquerque, J. S. de Sousa, G. A. Farias, Excitons in Si$_{1-x}$Ge$_x$ nanocrystals: Ab initio calculations, J. Appl. Phys. \textbf{103}, 103716 (2008).


\bibitem{thean2001} A. V. Thean, J.-P. Leburton, Stark effect and single-electron charging in silicon nanocrystal quantum dots, J. Appl. Phys. \textbf{89}, 2808 (2001).

\bibitem{he2005} L. He, A. Zunger, Multiple charging of InAs/GaAs quantum dots by electrons or holes: Addition energies and ground-state configurations, Phys. Rev. B \textbf{73}, 115324 (2006).

\bibitem{an2007} J. M. An, A. Franceschetti, A. Zunger, Electron and hole addition energies in PbSe quantum dots, Phys. Rev. B \textbf{76}, 045401 (2007).

\bibitem{szabo1996} A. Szabo, N. S. Ostlund, \emph{Modern Quantum Chemistry: Introduction to Advanced Electronic Structure Theory}, Dover, 1996.

\bibitem{soler96} P. Ordej\'on, E. Artacho, J. M . Soler, Self-consistent order-N density-functional calculations for very large systems, Phys. Rev. B \textbf{53}, R10441 (1996).

\bibitem{soler2002} J. M. Soler, E. Artacho, J. D. Gale, A. Garcia, J. Junquera, P. Ordej\'on, D. Sanchez-Portal, The SIESTA method for ab initio order-N materials simulation, J. Phys.: Condens.Matter. \textbf{14}, 2745 (2002).

\bibitem{ameen2016} T. A. Ameen, H. Ilatikhameneh, G. Klimeck, R. Rahman, Few-layer Phosphorene: An Ideal 2D Material For Tunnel Transistors, Sci. Rep. \textbf{6}, 28515 (2016).

\bibitem{kronik2012} L. Kronik,  T. Stein, S. Refaely-Abramson, R. Baer, Excitation Gaps of Finite-Sized Systems from Optimally Tuned Range-Separated Hybrid Functionals, J. Chem. Theory Comput. \textbf{8}, 1515 (2012). 
 
\end{thebibliography}
\end{document}